\documentclass[[journal]{IEEEtran}
\usepackage{cite}
\usepackage{amsmath,amssymb,amsfonts}
\usepackage{algorithmic}
\usepackage{graphicx}
\usepackage{textcomp}
\usepackage{float}
\usepackage{subfloat}
\usepackage{placeins}
\usepackage{caption}
\usepackage{subcaption}

\begin{document}
\title{Near-field Image Transmission and EVM Measurements in Rich Scattering Environment}
\author{Mir~Lodro, Gabriele~Gradoni , Christopher~Smartt, Ana~Vukovic, David~Thomas and Steve~Greedy  \vspace{-0.75cm}
\thanks{Mir Lodro, Steve Greedy, Christopher Smartt, Ana Vukovic, David Thomas and Gabriele Gradoni are with George Green Institute for Electromagnetic Research-GGIEMR, the University of Nottingham, UK. Gabriele Gradoni is also with British Telecommunications and the University of Cambridge, UK.}}
\maketitle
\begin{abstract}
In this work we present near-field image transmission and error vector magnitude measurement in rich scattering environment in metal enclosure. We check the effect of loading metal enclosure on the performance of SDR based near-field communication link. We focus on the key communication receiver parameters to observe the effect of near-field link in presence of rich-scattering and in presence of loading with RF absorber cones. The near-field performance is measured by transmitting wideband OFDM-modulated packets containing image information. Our finding suggest that the performance of OFDM based wideband near-field communication improves when metal enclosure is loaded with RF absorbers. Near-field EVM improves when the enclosure is loaded with RF absorber cones. Loading of the metal enclosure has effect of increased coherence bandwidth. Frequency selectivity was observed in an empty enclosure which suggests coherence bandwidth less than the signal bandwidth.  
\end{abstract}
\vspace{-0.17cm}

\vspace{-0.35cm}
\section{Introduction}
\label{section label}
Data transmission in the near-field in rich scattering environment for applications like wireless chip-to-chip communication and wireless network-on-chip (WNoC) communication has become important\cite{shamim2016wireless,chen2007inter,lodro2020near}. Near-field communication is traditionally used for low-rate and short-range communication that takes place using magnetic induction (MI) communication links.Magnetic induction communication depends on near-field component of magnetic dipole which decays as $1/r^3$ from the dipole. Hence, it may be suitable for low-rate information and power transfer\cite{kim2017review}. It has also been used for wireless underground sensor networks \cite{sun2010magnetic,kisseleff2018survey} and underwater communication \cite{akyildiz2015realizing,guo2017multiple}.Because of limited range it may offer frequency reuse and secure communication link. Hence, the receiver design for MI communication should be simple as there are no interfering components involved. However, it's not suitable technology for a scalable and high-data rate applications like wireless chip-to-chip communication. Additionally, the performance of MI links are subject to ambient effects and any misalignment between transmitter and receiver degrades the performance.Authors in \cite{kim2016near} have shown near-field magnetically coupled communication using loop antennas, however the performance of real-world communication receiver in the near-field is not investigated. Authors in \cite{shin2019rf} have proposed inductively coupled RF transceiver for wireless-chip-chip communication, however, the work has not introduced any signal quality based and end-to-end system performance analysis. There are few more studies on metal enclosure at high frequencies for wireless chip-to-chip communication such as \cite{fu2020modeling,timoneda2018channel,fu2019thz} but are limited to transfer function measurements and digital receiver design part is not investigated.The complexity of communication receiver design in the near-field has been high-lighted by author in \cite{mikki2020theory} and we link near-field effects on the communication signal quality. We have organized our paper in five sections. Section I is introduction which highlights the importance of near-field image transmission and EVM measurements. Section II is about packet structure and the receiver design stages for image transmission in near-field. Section III is PlutoSDR based near-field measurement setup in metal enclosure. Section IV is about experimental results and related discussions. This section summarizes results which includes channel estimation per OFDM symbols, constellation diagram for different modulation and coding rates, EVM measurements.

\vspace{-0.35cm}
\section{Packet Structure and Receiver Stages}
\begin{figure}
    \centerline{\includegraphics[width=\columnwidth]{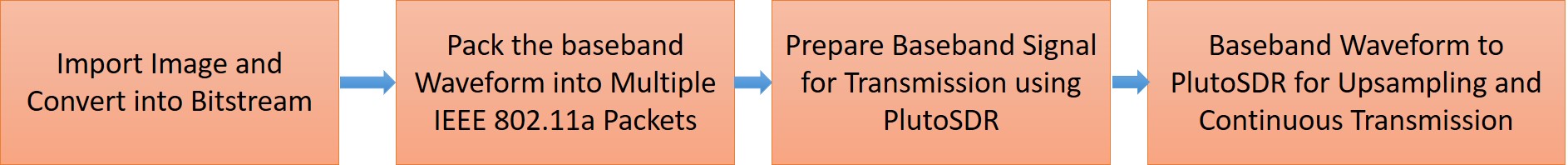}}
    \caption{Transmitter structure.}
    \label{fig:tx_flow}
\end{figure}

\begin{figure}
    \centerline{\includegraphics[width=\columnwidth]{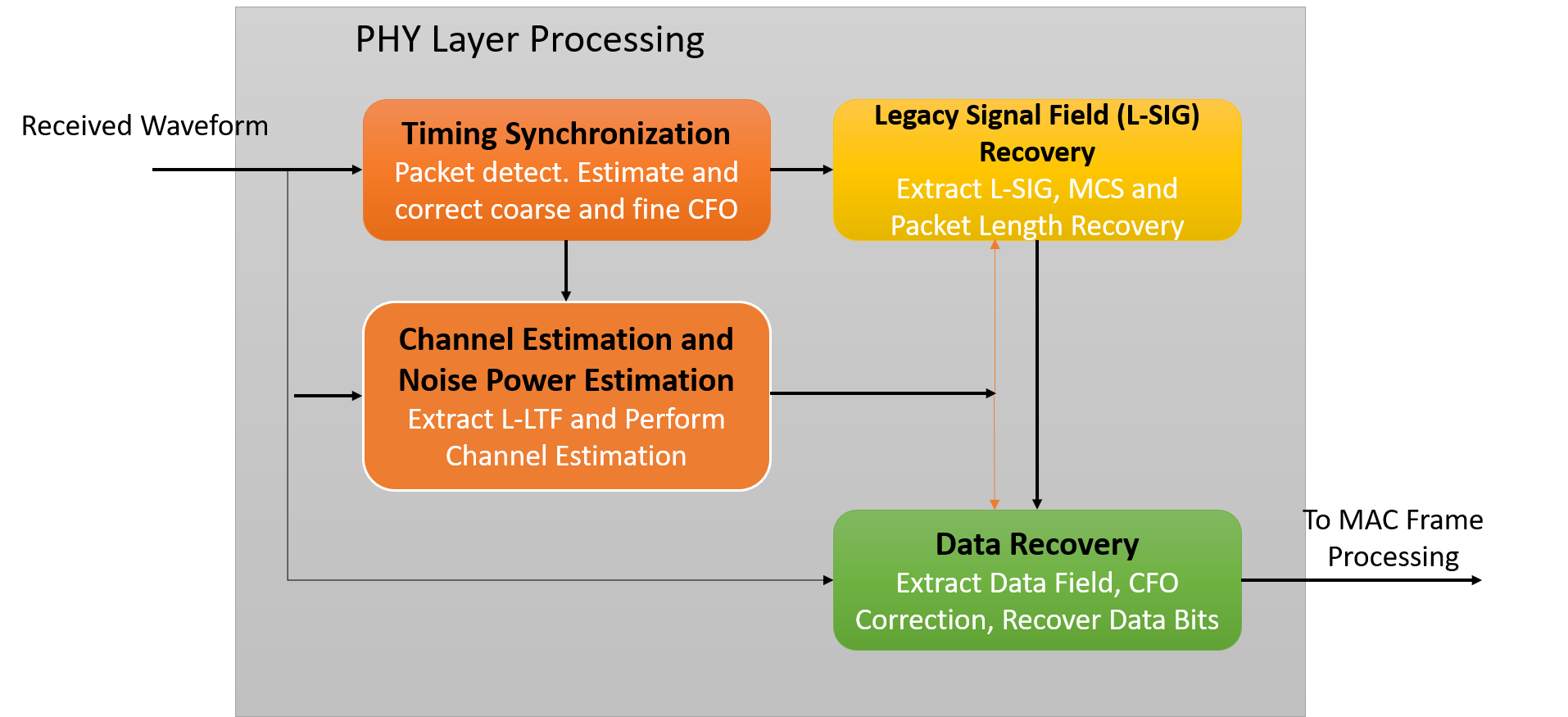}}
    \caption{PHY layer processing of received baseband waveform.}
    \label{fig:PHY_layer}
\end{figure}

An image file is encoded and converted into packets for over-the-air transmission in metal enclosure. Packet is divided into header fields and the data fields. There are three main fields which are used for channel estimation and information decoding. L-STF (Legacy-Short Training Field), L-LTF (Legacy-Long Training Field) and L-SIG (Legacy-Signal Field).Fig.~\ref{fig:tx_flow} shows simplified baseband transmitter structure where input image is first converted into bitstream and packed into IEEE 802.11a packets for upsampling and continuous transmission over the air using PlutoSDR. PlutoSDR has factory-default RF coverage from 325 MHz to 3.8 GHz and can be software upgraded from 70 MHz to 6.0GHz and configurable sample rate  from 520 kHz to 61.44 MHz. It has 12-bit ADC and DAC and can provide instantaneous bandwidth of 20 MHz. PlutoSDR is a full-duplex radio based on Analog Devices AD9363 which is highly integrated RF agile transceiver and Xilinx® Zynq Z-7010 FPGA.PlutoSDR can be interface with high using USB connection with host PC.Baseband waveform and other signal processing functionality can be performed using GNU Radio and MATLAB/Simulink. It's cost-effective full-duplex radio with high sampling rates and instantaneous bandwidth.Fig.~\ref{fig:PHY_layer} shows PHY layer baseband processing of received baseband waveform. First step is to perform packet detection and in later steps perform coarse frequency offset estimation and correction, fine frequency offset estimation and correction. Thereafter, channel estimation and noise variance estimation is performed using L-LTF then L-SIG and MCS recovery is performed whose output is supplied to data recovery stage where data bits are recovered.

\vspace{-0.35cm}
\section{Measurement Setup}
\label{sec:measurement_setup}
\begin{figure}
\centerline{\includegraphics[width=0.7\linewidth]{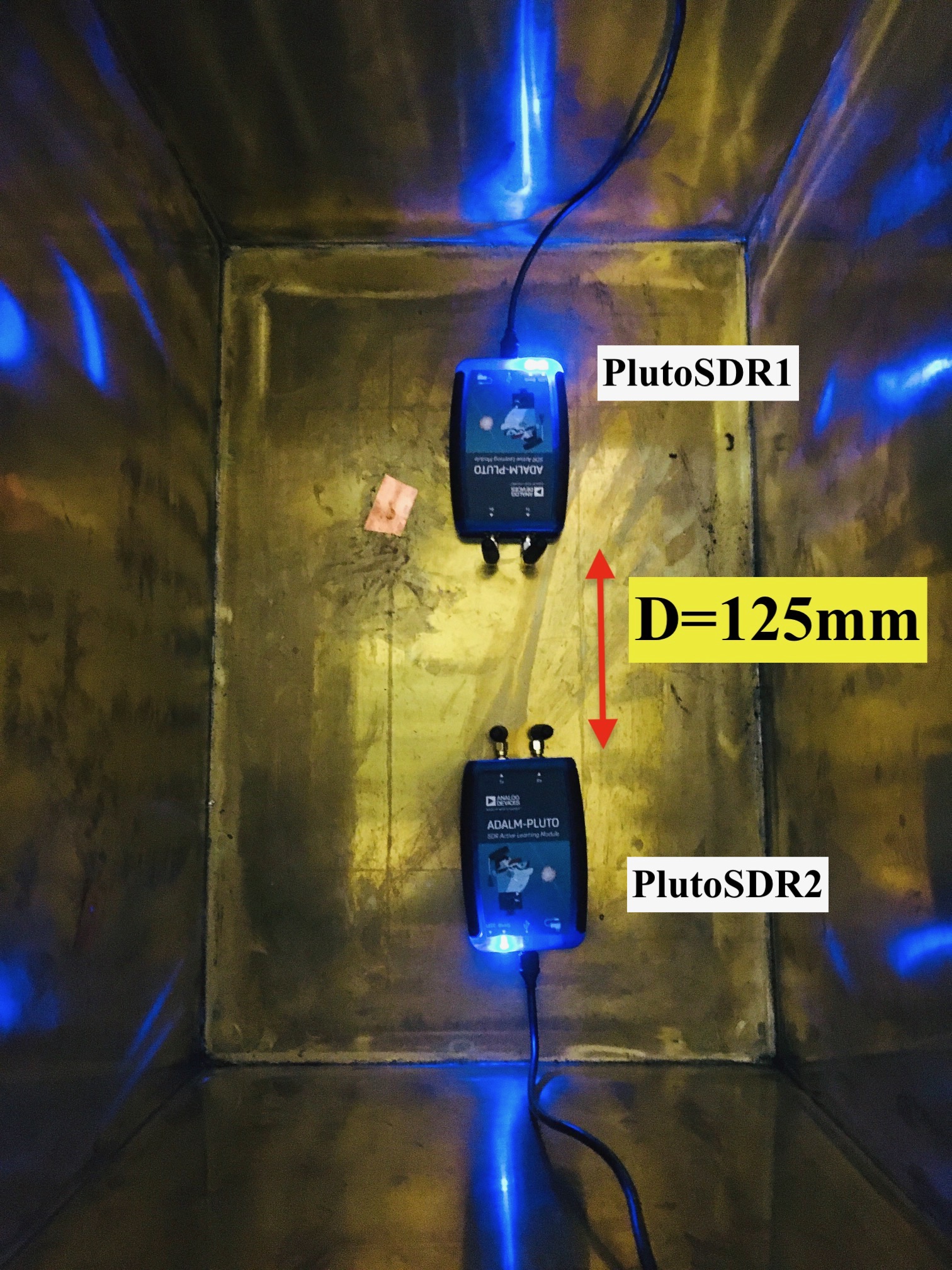}}
\caption{PlutoSDR based measurement setup with communication link distance $D=125\,mm$.}
\label{fig:far-field}
\end{figure}

\begin{figure}[h]
    \centering
    \subfloat[RF absorbers in the middle left side]{\label{fig:center}\includegraphics[width=0.6\linewidth,angle=90]{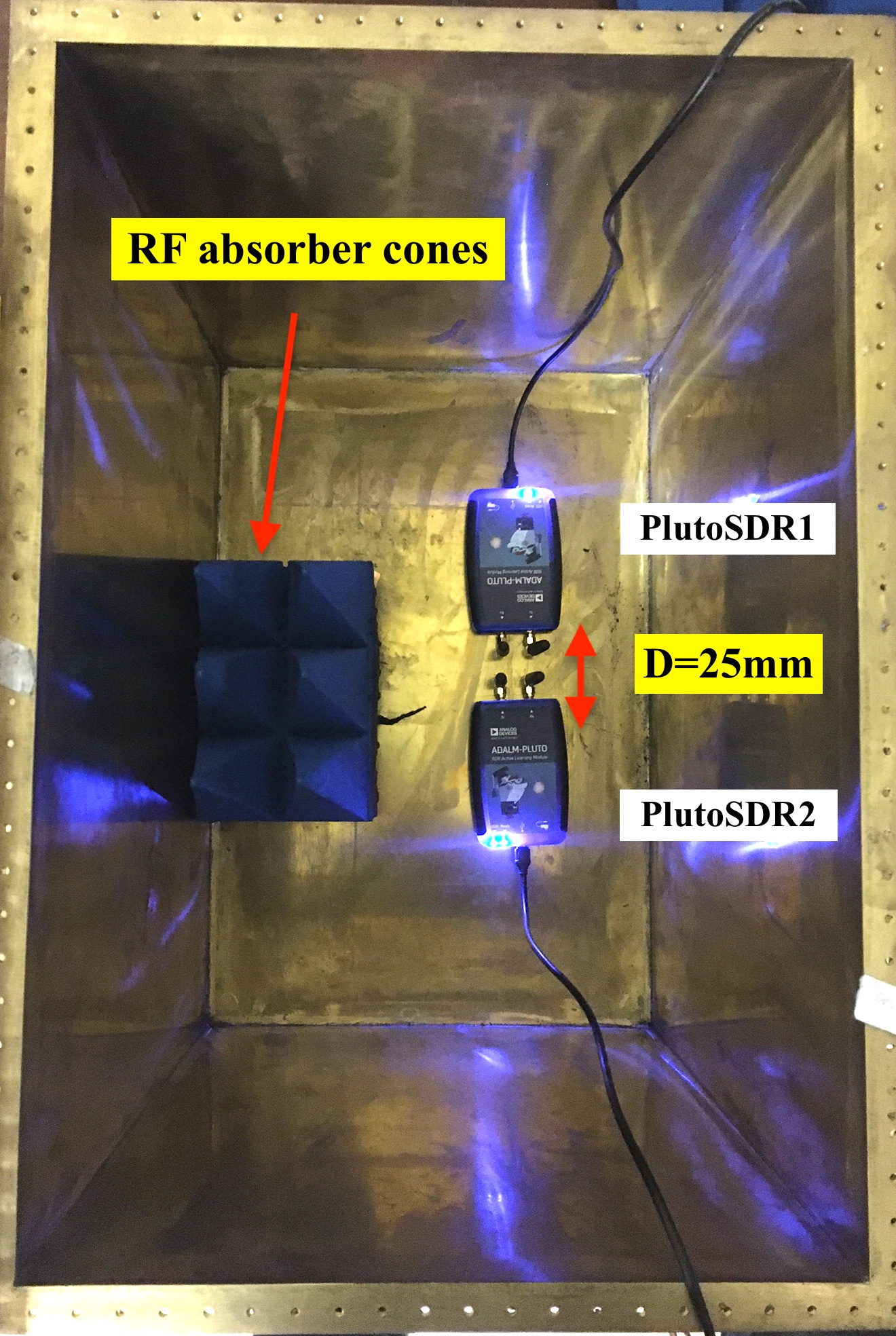}}\hspace{1ex}
    \subfloat[RF absorbers in the corner]{\label{fig:corner}\includegraphics[width=0.6\linewidth,angle=90]{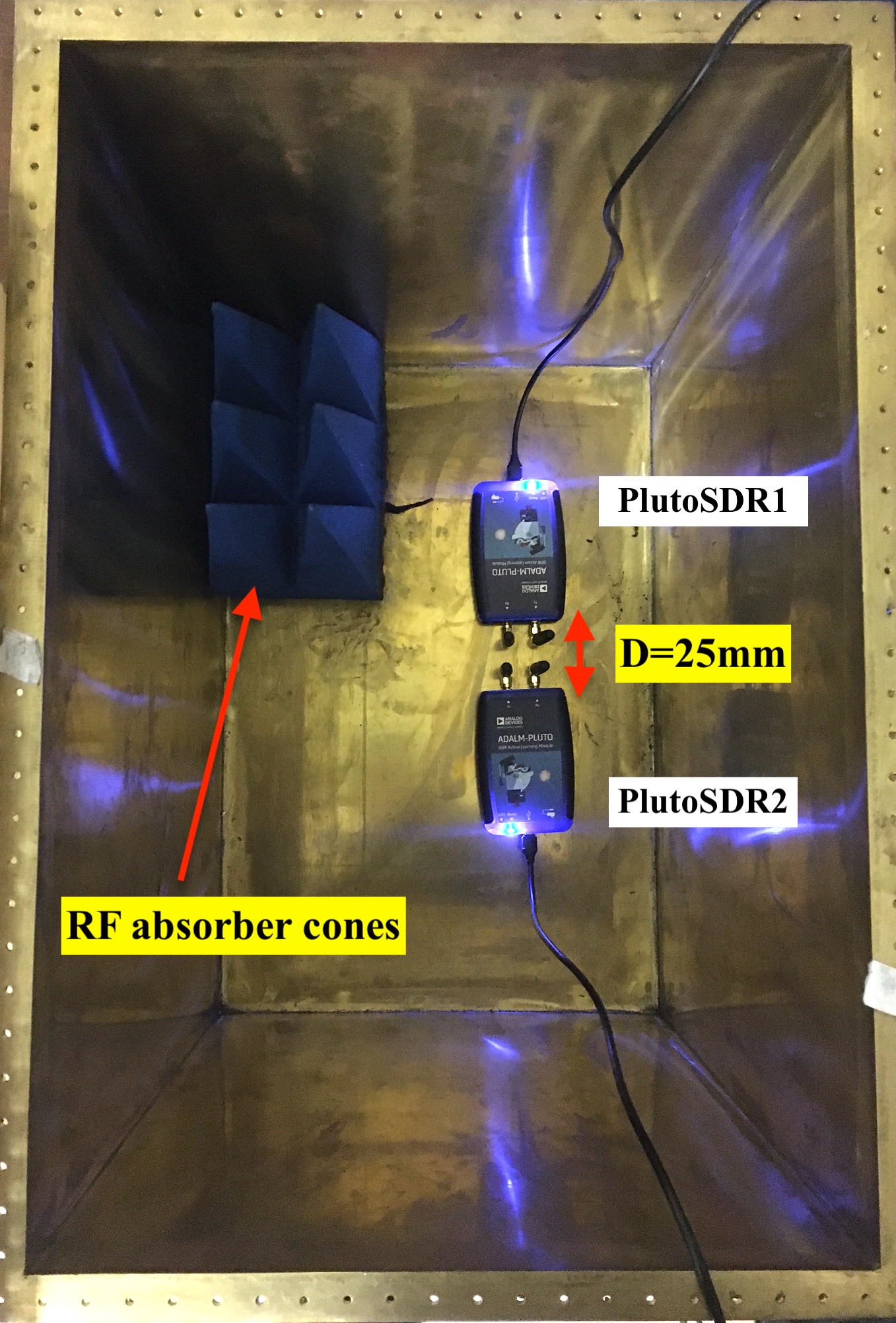}}
    \caption{PlutoSDR based near-field measurement setup in metal enclosure with RF absorber cones in middle left side and corner respectively.}
    \label{fig:near_field}
\end{figure}

We have used a brass metal enclosure with dimensions of $h\times \ell \times w$ of $45\,cm\times 37\,cm\times 55\,cm$.The metal enclosure is a perfectly reflecting environment and when excited at 2.432 GHz it produces rich isotropic multipath fading environment. Two plutoSDRs are connected to same Intel Xeon host PC.
The baseband waveform is upconverted for over-the-air transmission using PlutoSDR for repeated transmission functionality. The received waveform is captured using another PlutoSDR and baseband waveform is decoded for further information decoding and visualization. An image file is fragmented into multiple MAC service data units (MSDU). MAC protocol data unit (MPDU) is created from each MSDU. MPDU is further passed to PHY layer and PSDU are formed where each PSDU is packet into nonHT packet. The number of generated packets primarily depends on the MSDU length and the modulation and coding scheme selected. Multiple PSDU are processed to form PLCP protocol data unit (PPDU) which are upconverted for RF transmission. RF transmission is captured using another PlutoSDR and baseband samples are processed to convert extracted MPDUs into MSDUs. Information extracted from multiple MSDUs is synthesized to recover information bits and transmitted image is recovered.
Fig.~\ref{fig:far-field} shows radiative near-field measurement setup where two PlutoSDRs are separated by a distance of 125 mm. Fig.~\ref{fig:near_field} shows near-field measurement setup where two PlutoSDRs are separated by a distance of 25 mm.
Fig.~\ref{fig:near_field} shows near-field measurement setup in presence of RF absorbers at two locations inside the metal enclosure. There are six RF absorbers in middle left side of the enclosure as in Fig.~\ref{fig:center} just next to near-field communication link and the same number of RF absorbers are placed in the corner of the enclosure.
Table~\ref{tab1} shows configuration parameters and the corresponding values. Tx and Rx communication takes place at 2.432 GHz which is channel 5 of 2.4GHz band. Non-HT packet idle time $20\mu s$.
\vspace{10mm}
\begin{table}[h!]
\renewcommand{\arraystretch}{1.3}
\caption{Parameters Configuration }
\vskip0.2in
\begin{center}
\small \begin{tabular}{|c|c|c|}\hline
Parameters & Values  \\ \hline
MSDU Length & 2304   \\ \hline
Modulation and Coding Techniques&MCS0-MCS7 \\\hline
Channel Bandwidth& 20MHz\\ \hline
Center Frequency&2.432 GHz\\ \hline
Non-HT Packet idle Time& $20\mu s$\\ \hline 
\end{tabular}
\end{center}
\label{tab1}
\end{table}

\vspace{-0.65cm}
\section{Experimental Results and Discussions}
We performed near-field measurements in the empty metal enclosure and two levels of loadings. Empty metal enclosure produces a rich isotropic multipath scattering environment. Loading of the enclosure reduces isotropic scattering and increases the coherence bandwidth of the wireless channel. It's important for the wideband communication link to avoid channel frequency-selectivity. We observed channel selectivity in case of an empty enclosure, however, the channel selectivity was reduced after introducing loss in the metal enclosure. We focused on RMS and peak EVM of BPSK, QPSK, and M-QAM modulation techniques with different coding rates. We visually observed inter-symbol interference in the case of the empty enclosure and this effect was removed by increasing the coherence bandwidth of the channel.

\begin{figure}
    \centering
    \subfloat[Captured data packets]{\label{fig:rxwaveform}\includegraphics[width=0.45\textwidth]{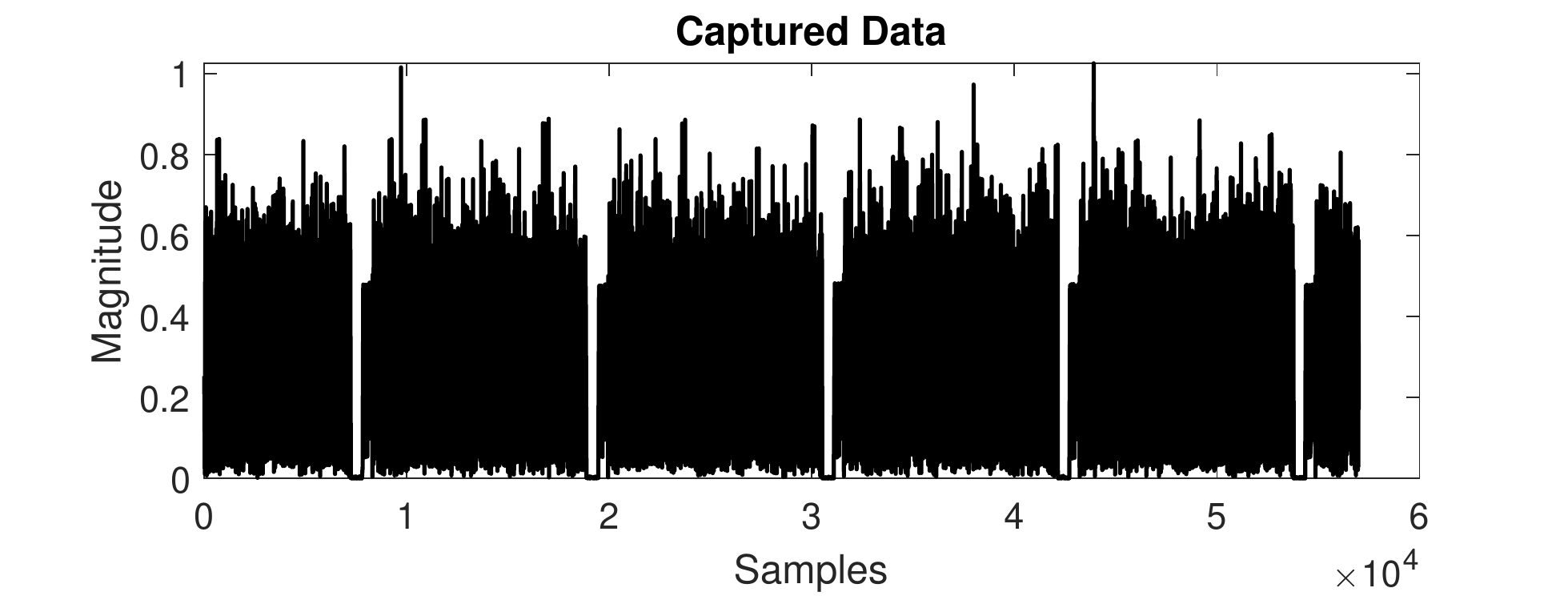}}\\
    \subfloat[Struckture of nonHT Packet]{\label{fig:notHT_packets}\includegraphics[width=0.45\textwidth]{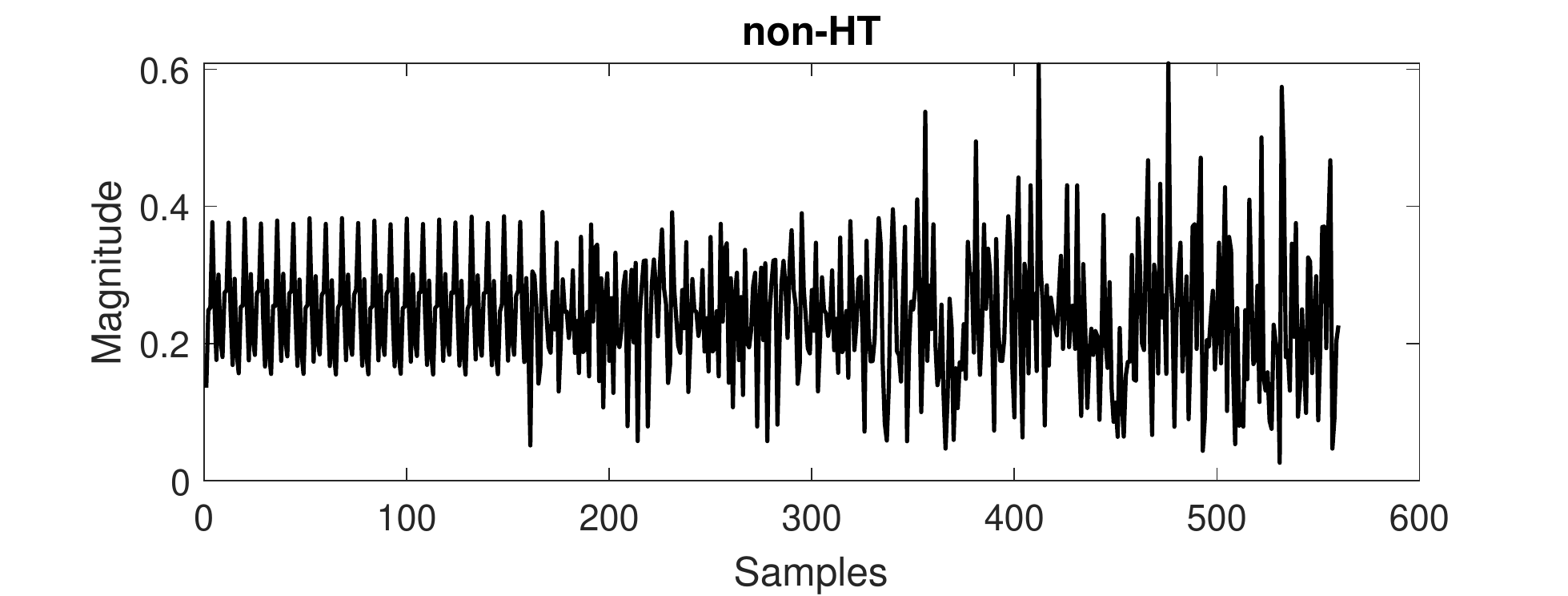}}
    \caption{Captured data packets and nonHT packet at Tx-Rx distance of 125mm.}
    \label{fig:data_packets_nonHT}
\end{figure}
Fig. \ref{fig:rxwaveform} shows a typical over-the-air captured data packet waveform with idle time of $20 \mu s$. Fig. \ref{fig:data_packets_nonHT} shows structure of received nonHT data packet.
\begin{figure}
    \centering
    \subfloat[64-QAM rate 3/4:empty]{\label{const_mcs7_125mm}\includegraphics[width=0.5\linewidth]{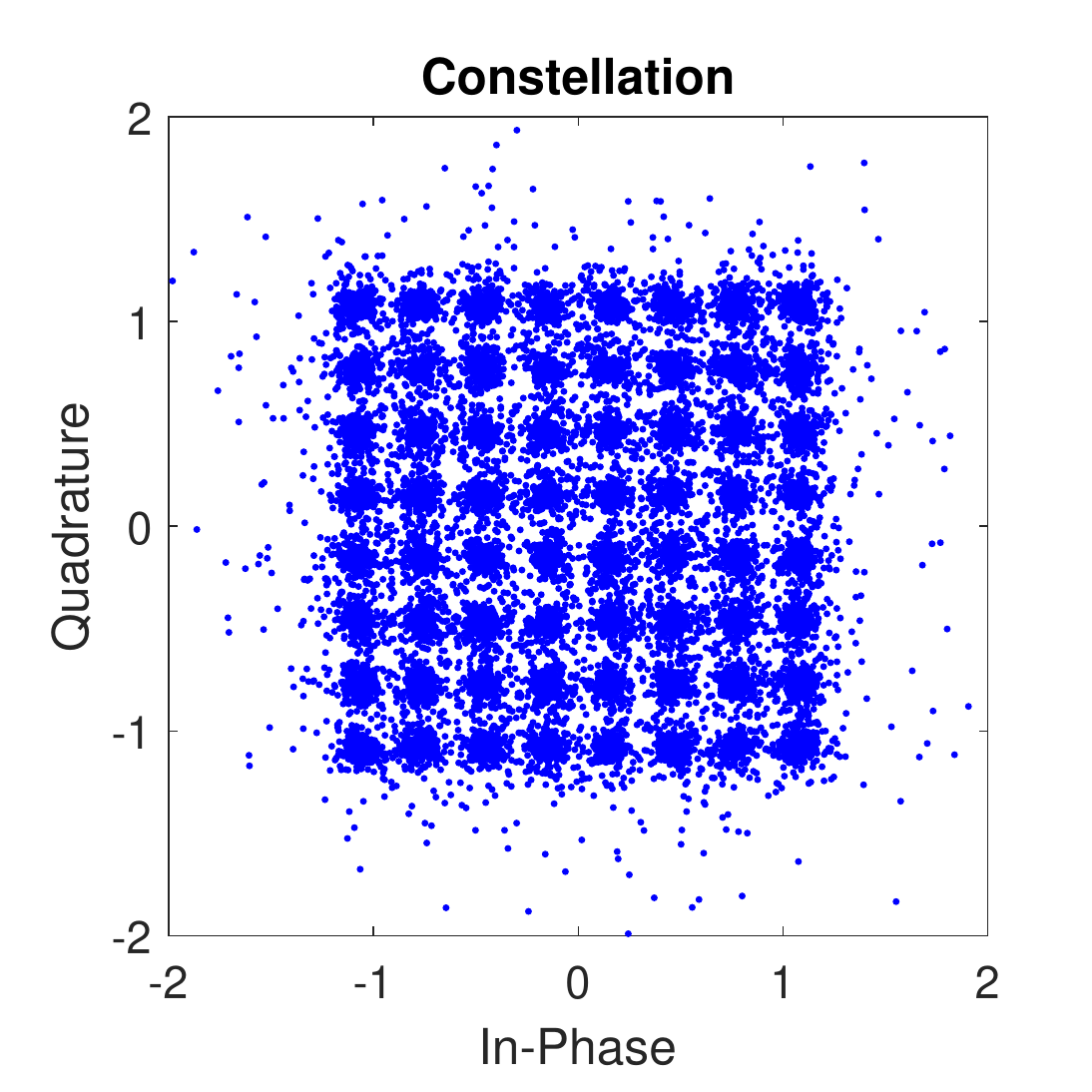}}
    \subfloat[64-QAM rate 3/4:losses]{\label{const_mcs7_6cones_125mm}\includegraphics[width=0.5\linewidth]{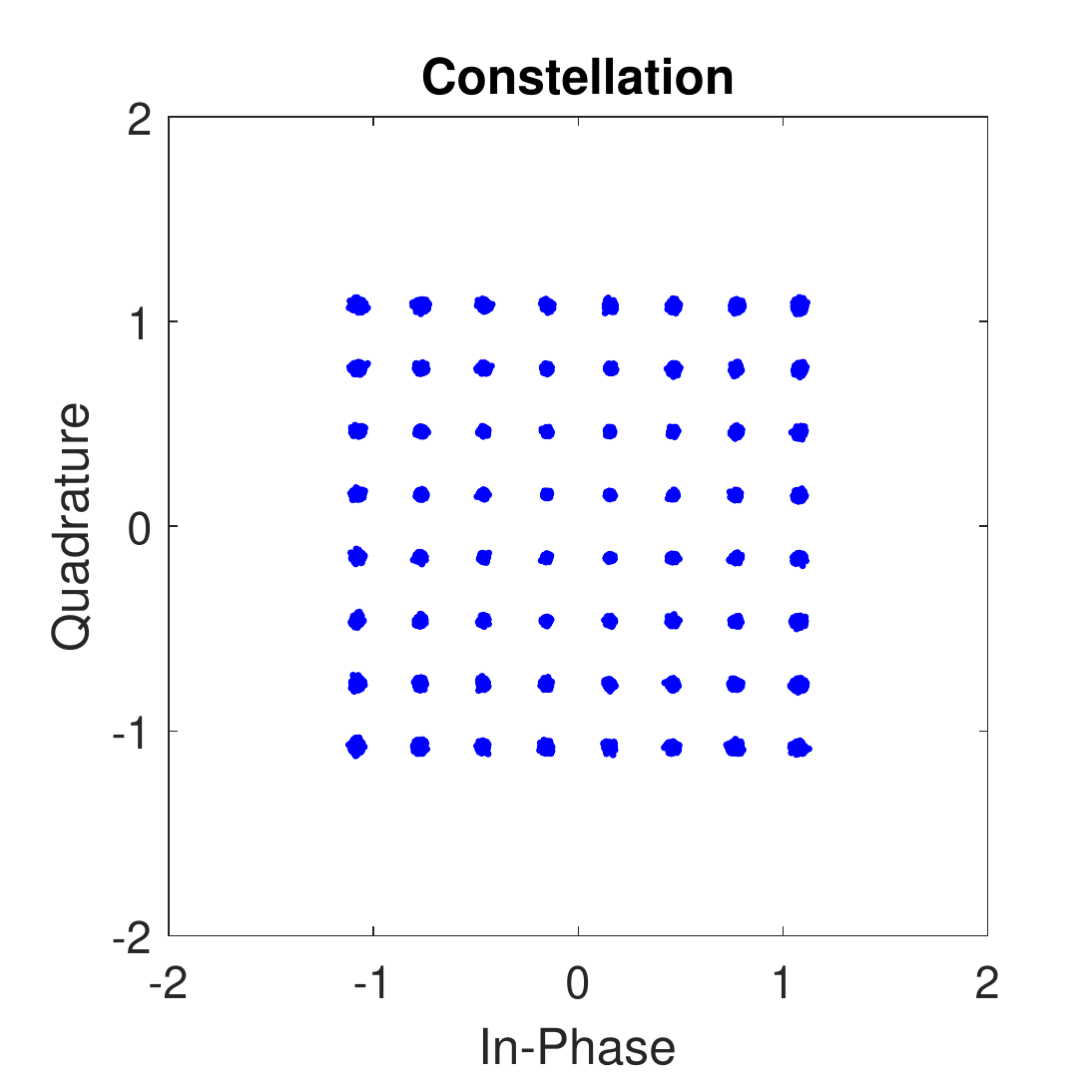}}
    \caption{Equalized constellation diagram of 64-QAM rate 3/4 at D=125 mm empty enclosure and when metal enclosure is loaded with losses.}
    \label{fig:const_125mm}
\end{figure}
Fig.~\ref{fig:const_125mm} shows the received constellation diagram of 64-QAM rate 3/4 at a distance of 125mm in the absence and presence of loading. It can be seen from the two constellation diagrams that there are many timing errors when the metal enclosure is empty and the timing errors are reduced when six RF absorber cones are introduced in the enclosure. 

\begin{figure}
    \centering
    \subfloat[]{\includegraphics[width=0.8\linewidth]{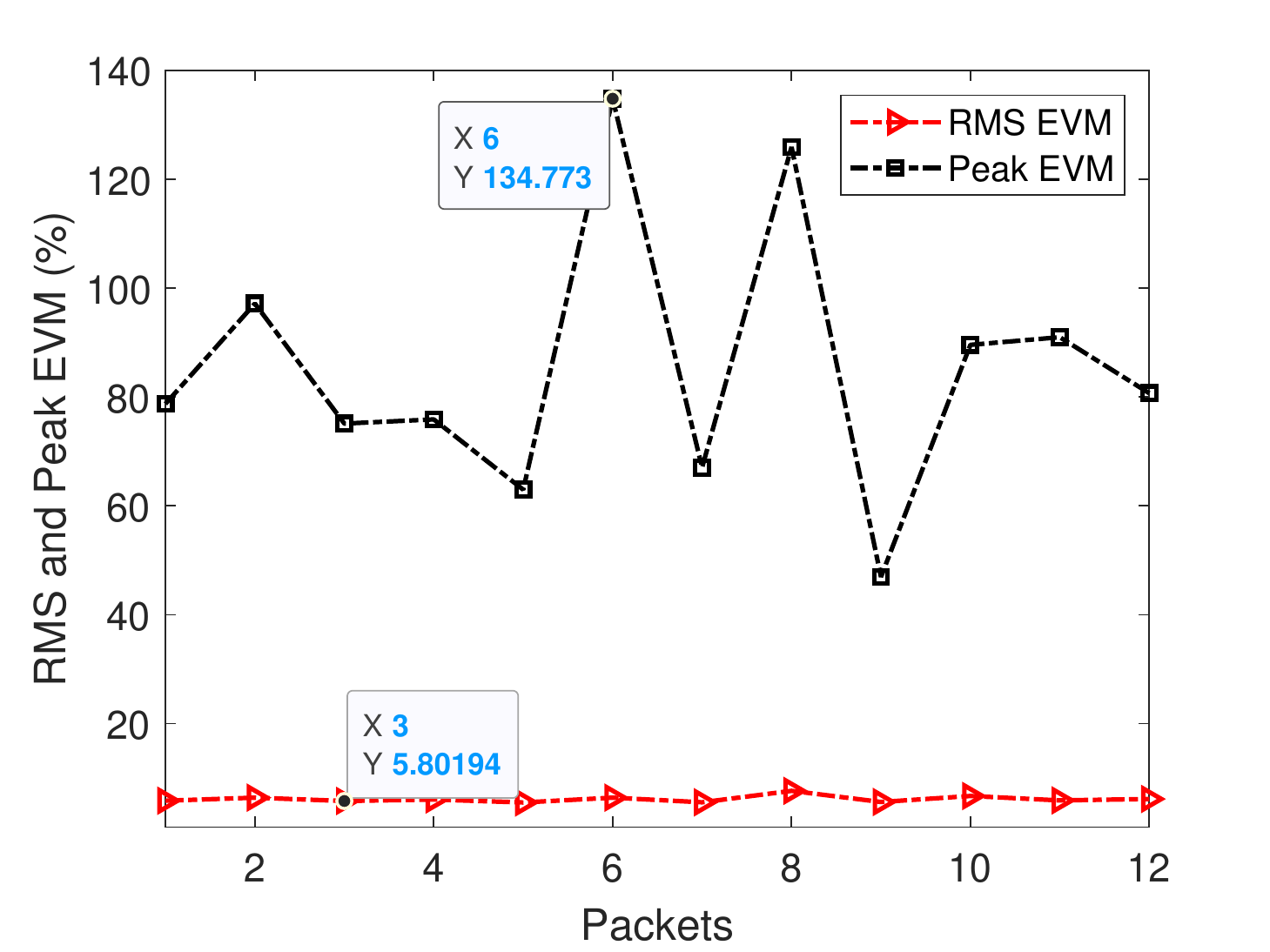}}\\
    \subfloat[]{\includegraphics[width=0.8\linewidth]{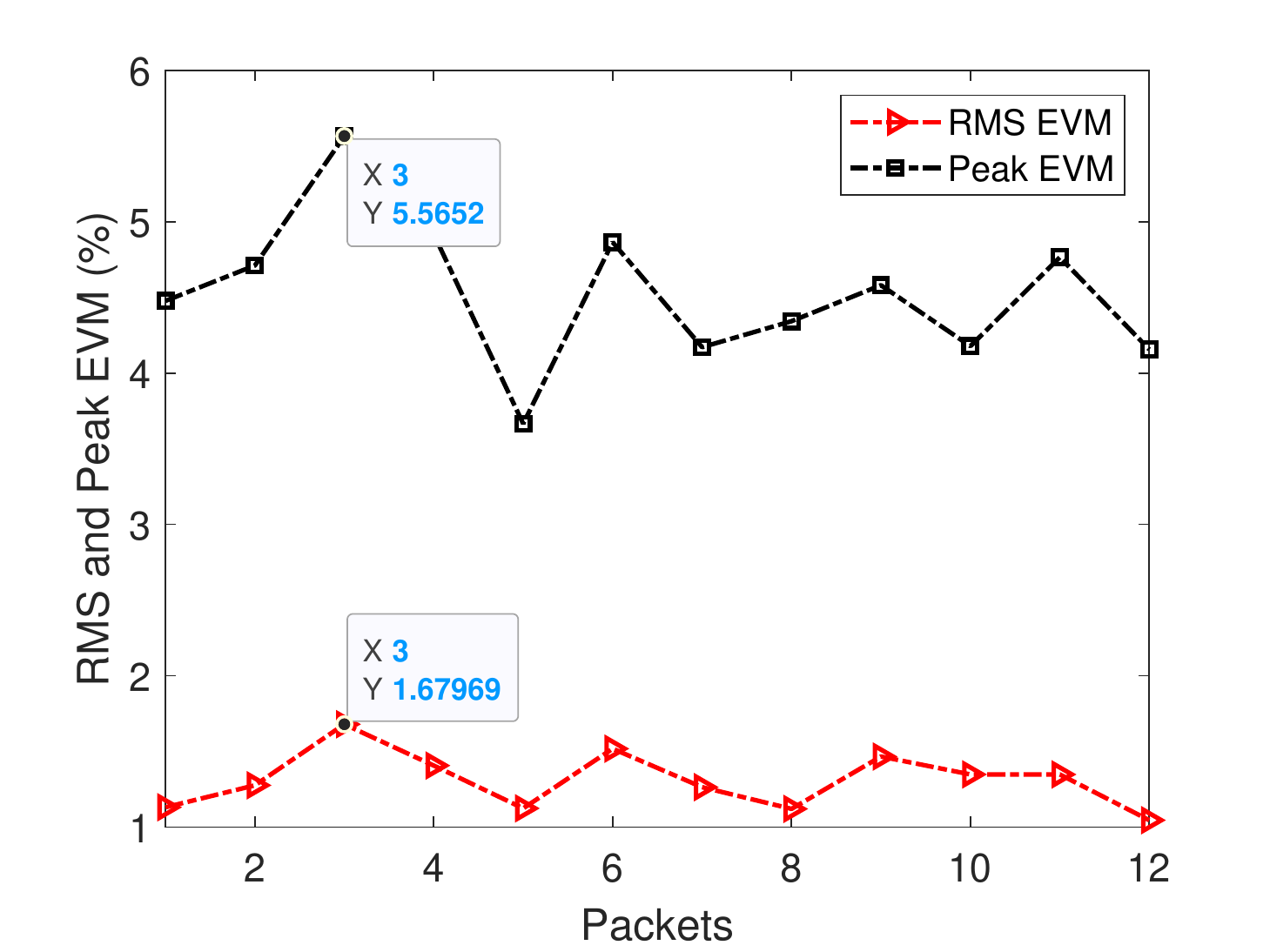}}
    \caption{RMS and Peak EVM of 64-QAM rate 3/4 at distance of 125mm in empty metal enclosure and when it was loaded with losses.}
    \label{fig:evm_empty_loaded}
\end{figure}
Fig.~\ref{fig:evm_empty_loaded} shows RMS and Peak EVM in a rich scattering environment and when the enclosure is loaded with six RF absorber cones. The peak EVM is $134.77\%$ in the case of the empty enclosure and with the addition of losses peak EVM is reduced to $5.56\%$. Similarly, RMS EVM has been reduced in presence of absorbers from $5.8\%$ to $1.67\%$.
\begin{figure}
    \centering
   \subfloat[]{\label{fig:chanEst_mcs7_125mm}\includegraphics[width=0.8\linewidth]{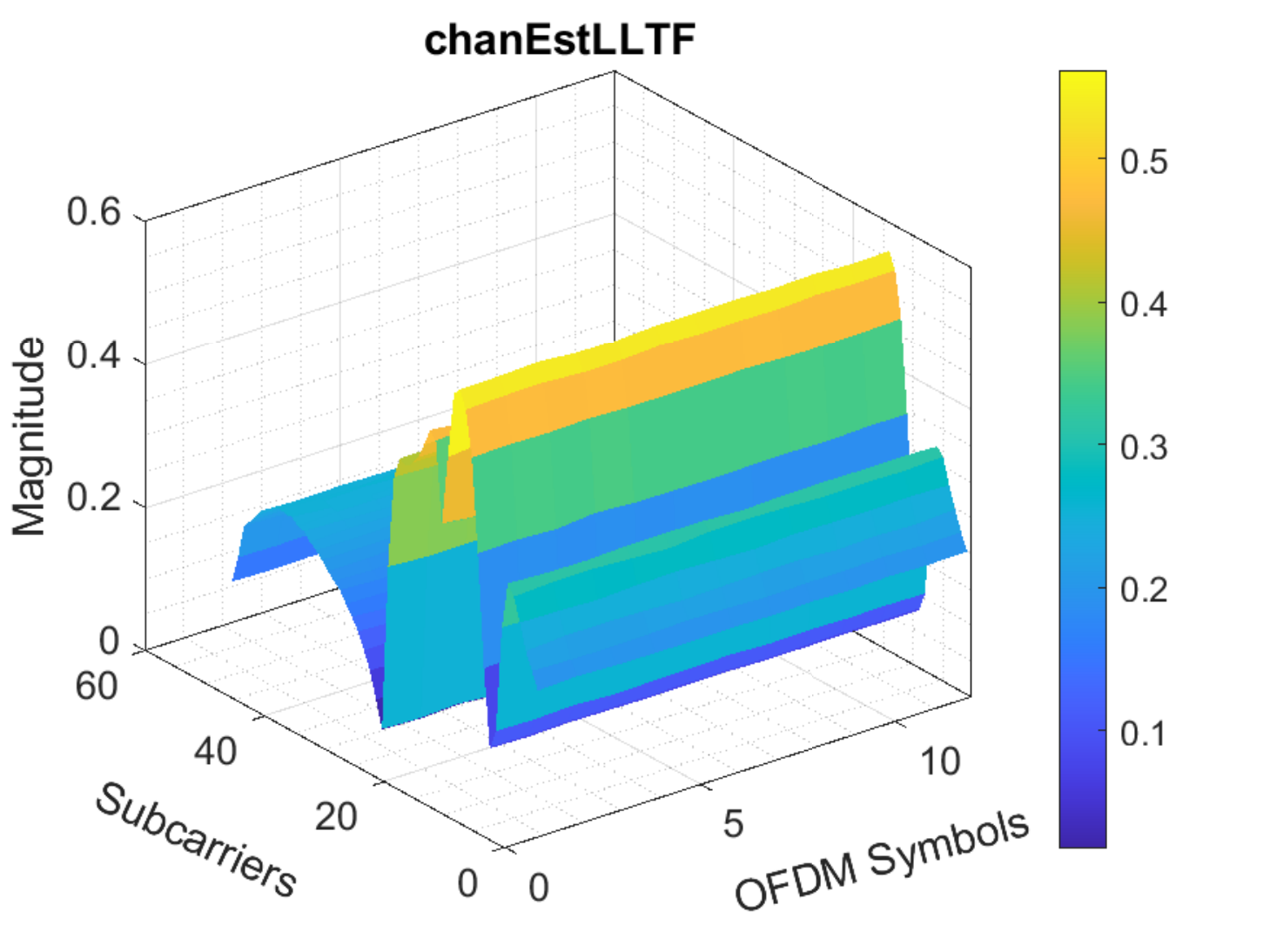}}\\
   \subfloat[]{\label{fig:chanEst_mcs7_6cones_125mm}\includegraphics[width=0.8\linewidth]{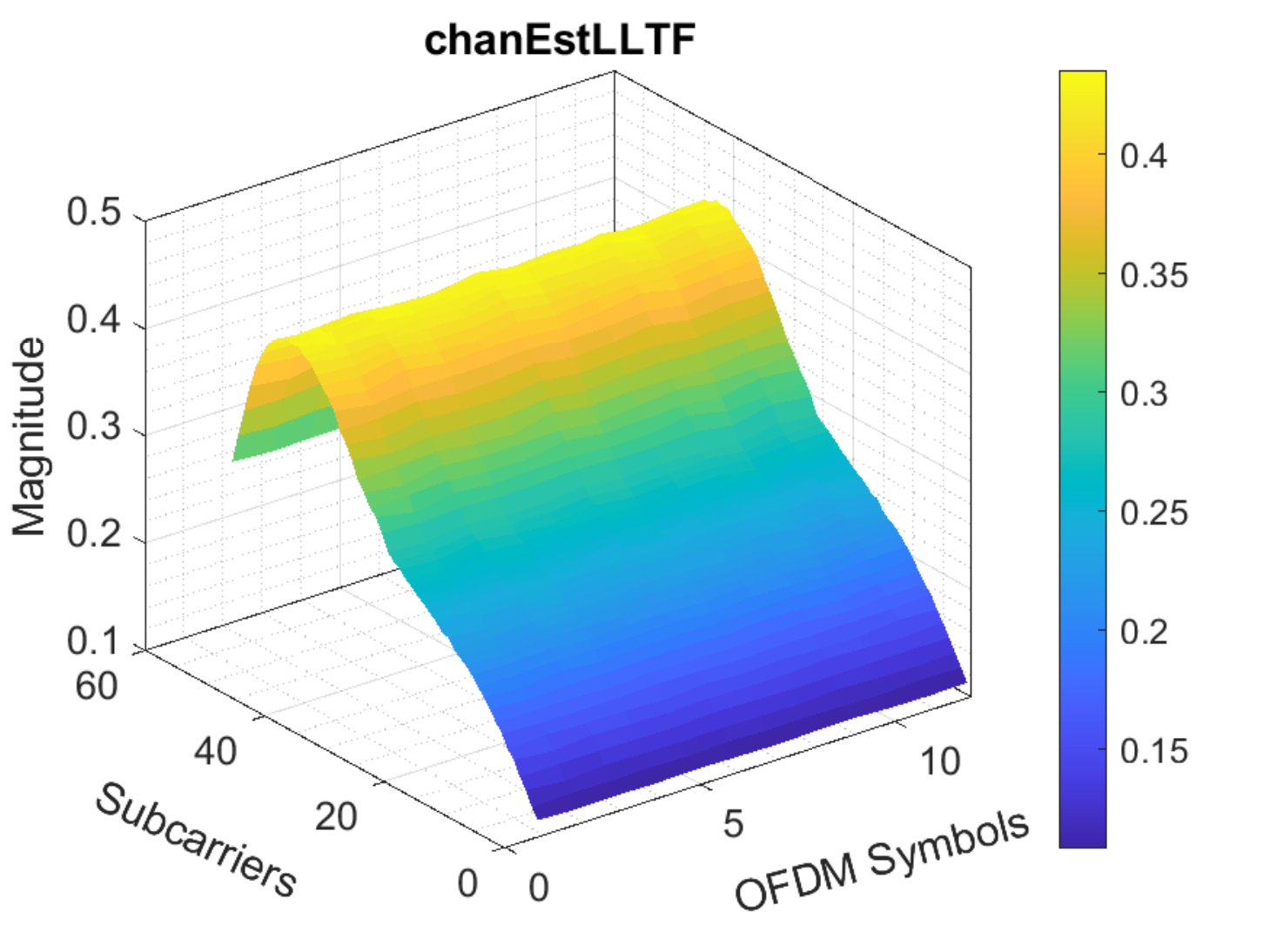}}
    \caption{Channel estimation at distance of 125mm when enclosure was empty and when it was loaded with losses.}
    \label{fig:chanEst_mcs7_125mm}
\end{figure}

\begin{figure}
    \centering
    \subfloat[]{\label{fig:const_mc7_25mm}\includegraphics[width=0.45\linewidth]{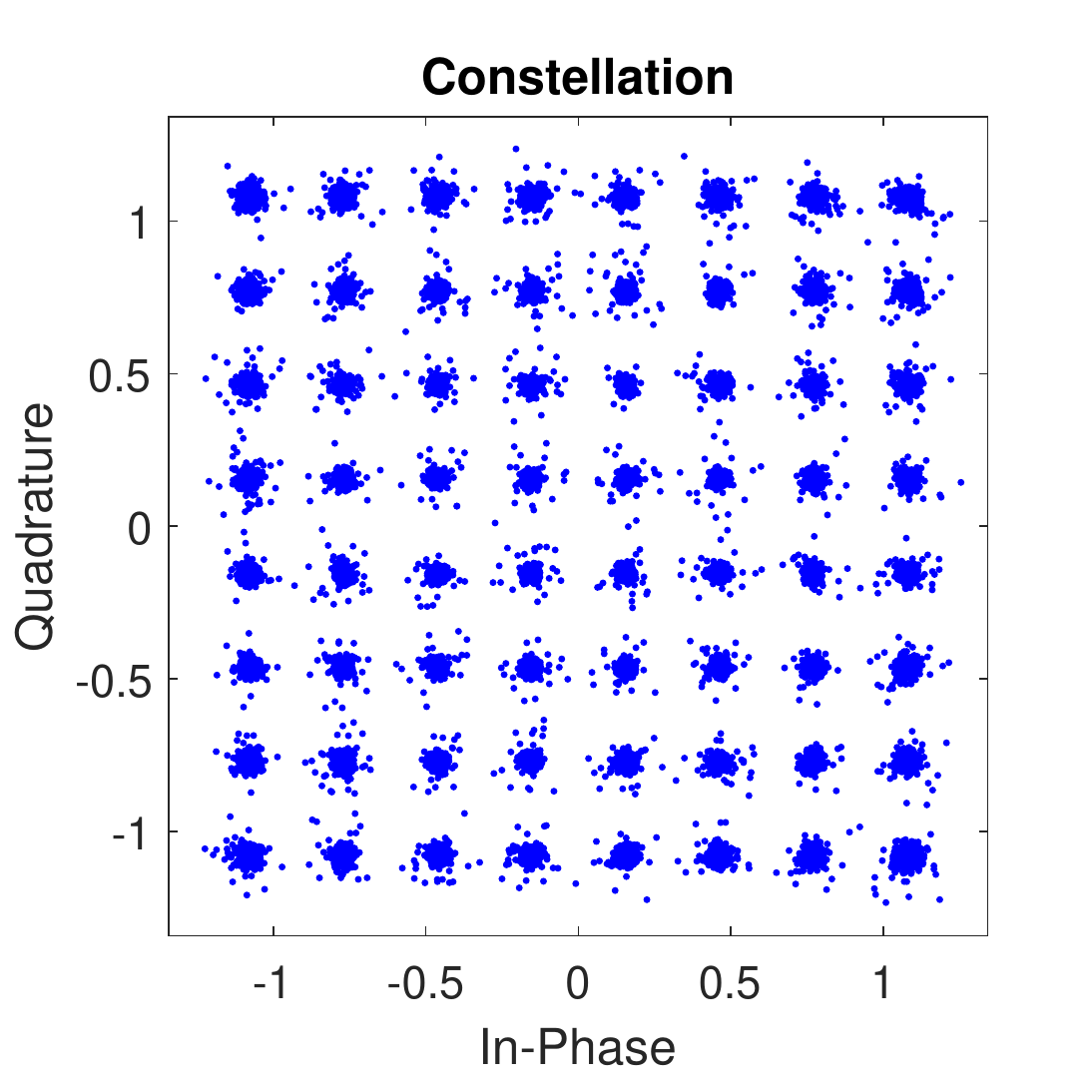}}
    \subfloat[]{\label{fig:const_mc7_6cones_25mm}\includegraphics[width=0.45\linewidth]{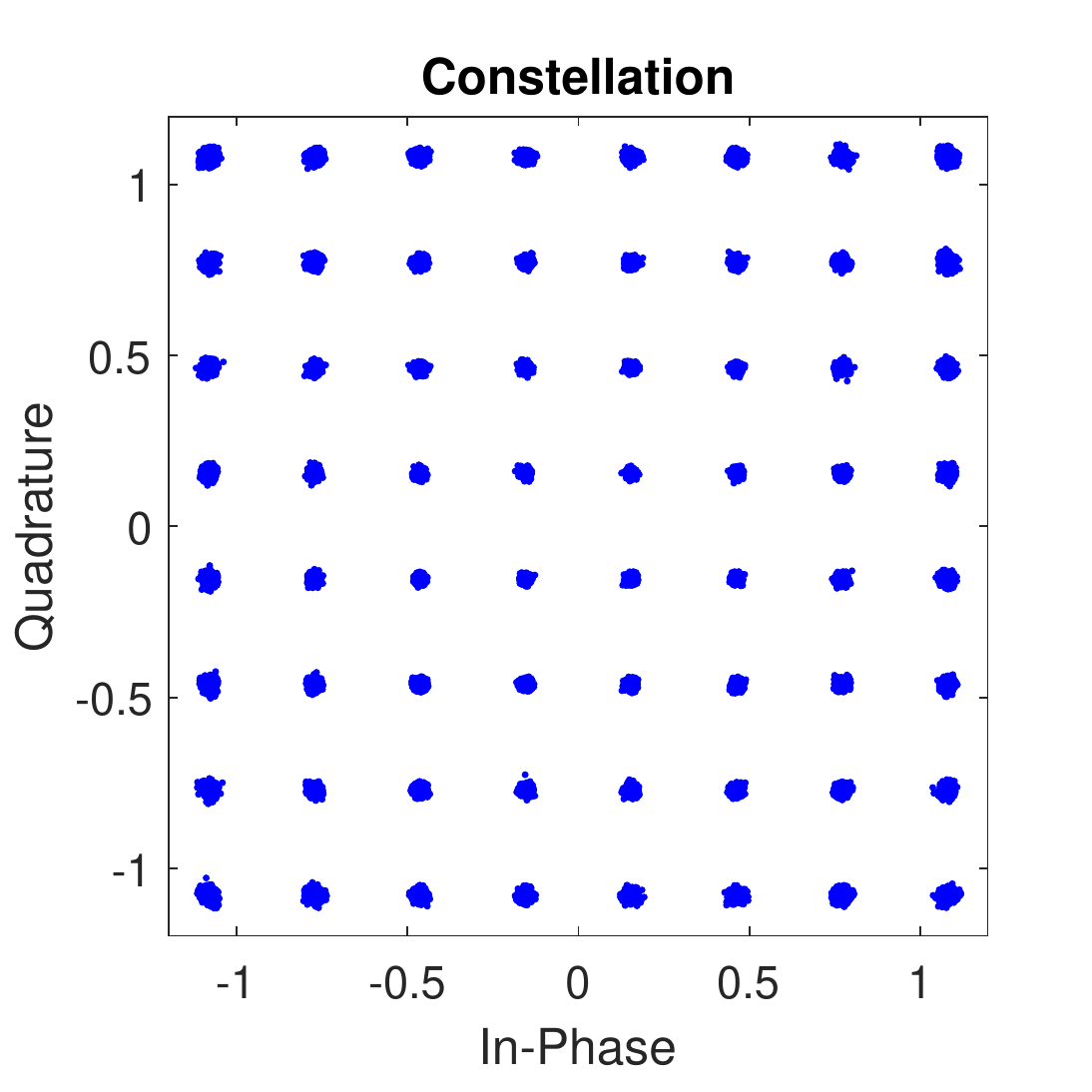}}\\
    \subfloat[]{\label{fig:const_mc7_corner_25mm}\includegraphics[width=0.45\linewidth]{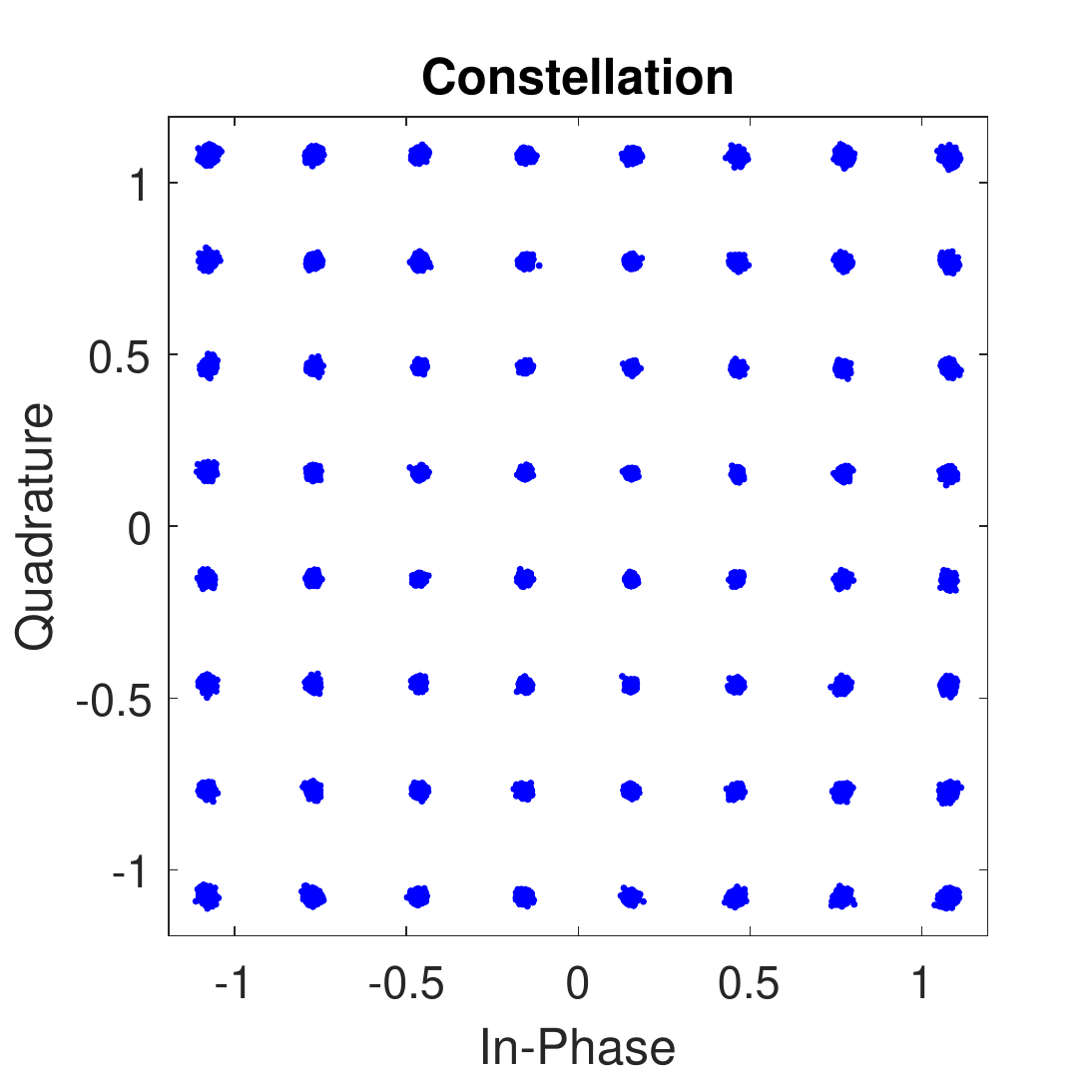}}
    \caption{Constellation diagram of 64-QAM 3/4 at distance of 25mm when the enclosure was empty and the loss was added in the middle left side and the corner of the metal enclosure.}
    \label{fig:const_25mm}
\end{figure}

Fig.~\ref{fig:chanEst_mcs7_125mm} shows channel state information for Tx-Rx distance of 125mm in the empty enclosure and in presence of loss. Fig.~\ref{fig:chanEst_mcs7_125mm} shows channel estimation in empty enclosure. It can be seen from the figure that there is frequency selectivity in the channel, which is addressed by placing the absorbers in the metal enclosure \cite{chen2009channel,chen2011estimation}. Frequency-selectivity in case of loaded enclosure as in Fig.~\ref{fig:chanEst_mcs7_6cones_125mm} has been reduced in contrast to empty enclosure.
Fig.~\ref{fig:const_25mm} shows constellation diagram of 64-QAM rate 3/4 for Tx-Rx distance of 25mm for empty enclosure and when the enclosure is loaded with RF absorbers at two locations inside metal enclosure. It can been seen from Fig.~\ref{fig:const_mc7_25mm} that signal quality is poor in contrast to signal quality in presence of loss as shown in Fig.\ref{fig:const_mc7_6cones_25mm} and Fig.\ref{fig:const_mc7_corner_25mm} respectively.

\begin{figure}
    \centering
    \subfloat[]{\includegraphics[width=0.8\linewidth]{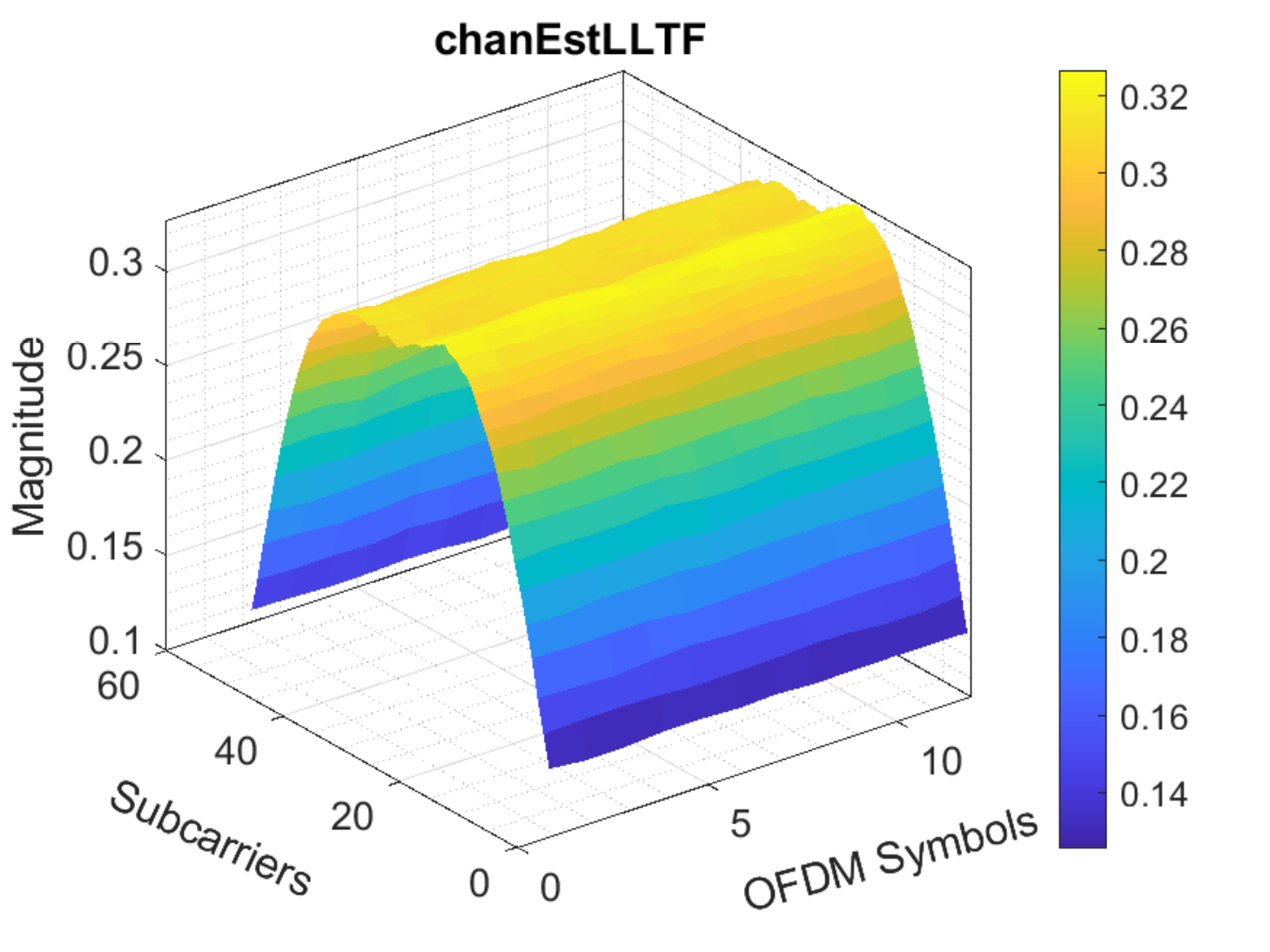}}\\
    \subfloat[]{\includegraphics[width=0.8\linewidth]{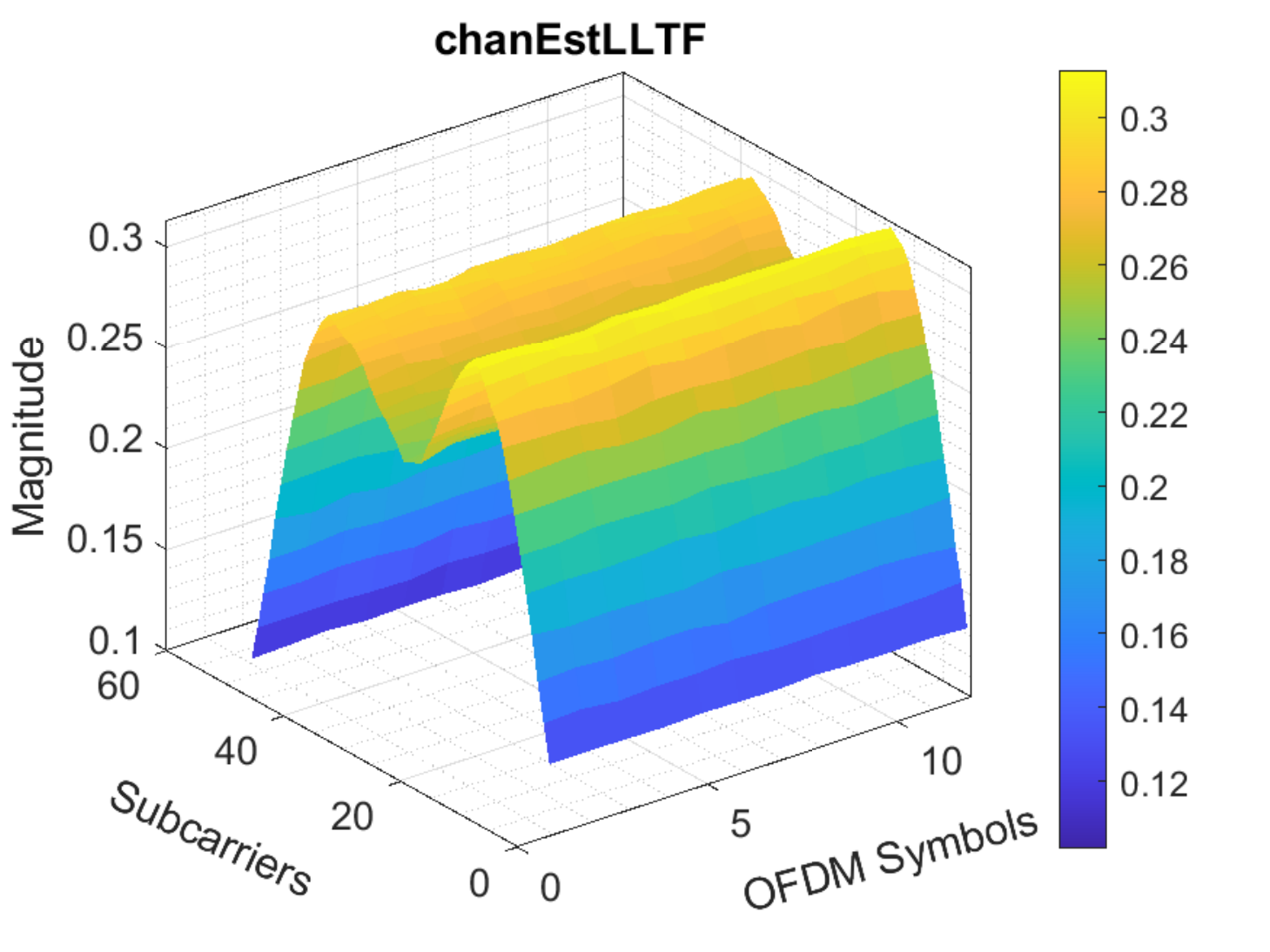}}
    \caption{Channel estimation at distance of 25mm when the RF absorber cones were placed at middle left and corner of the metal enclosure respectively.}
    \label{fig:chan_Est_25mm}
\end{figure}

\begin{figure}
    \centering
    \subfloat[]{\includegraphics[width=0.8\linewidth]{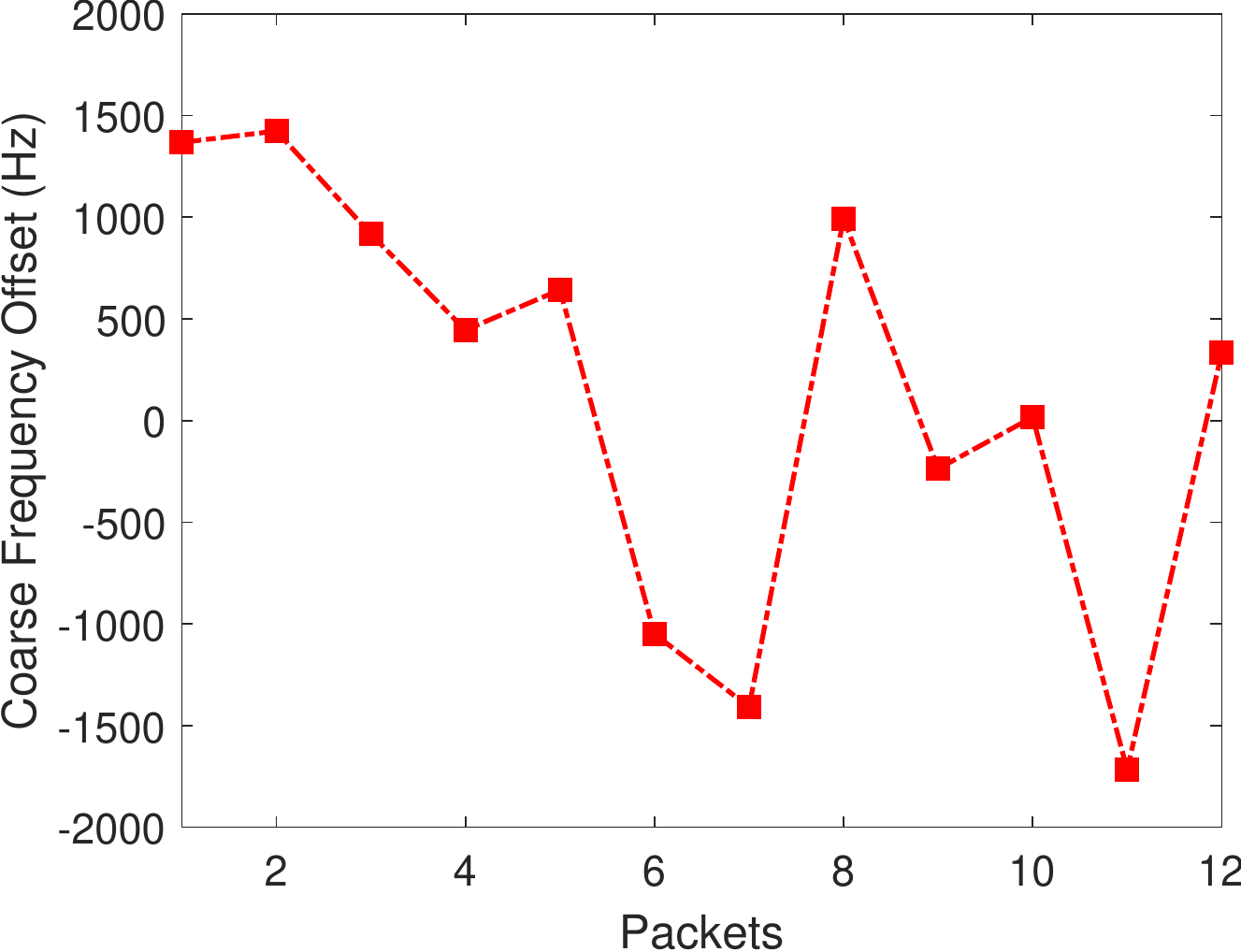}}\hspace{1ex}
    \subfloat[]{\label{fig:avg_cfo}\includegraphics[width=0.8\linewidth]{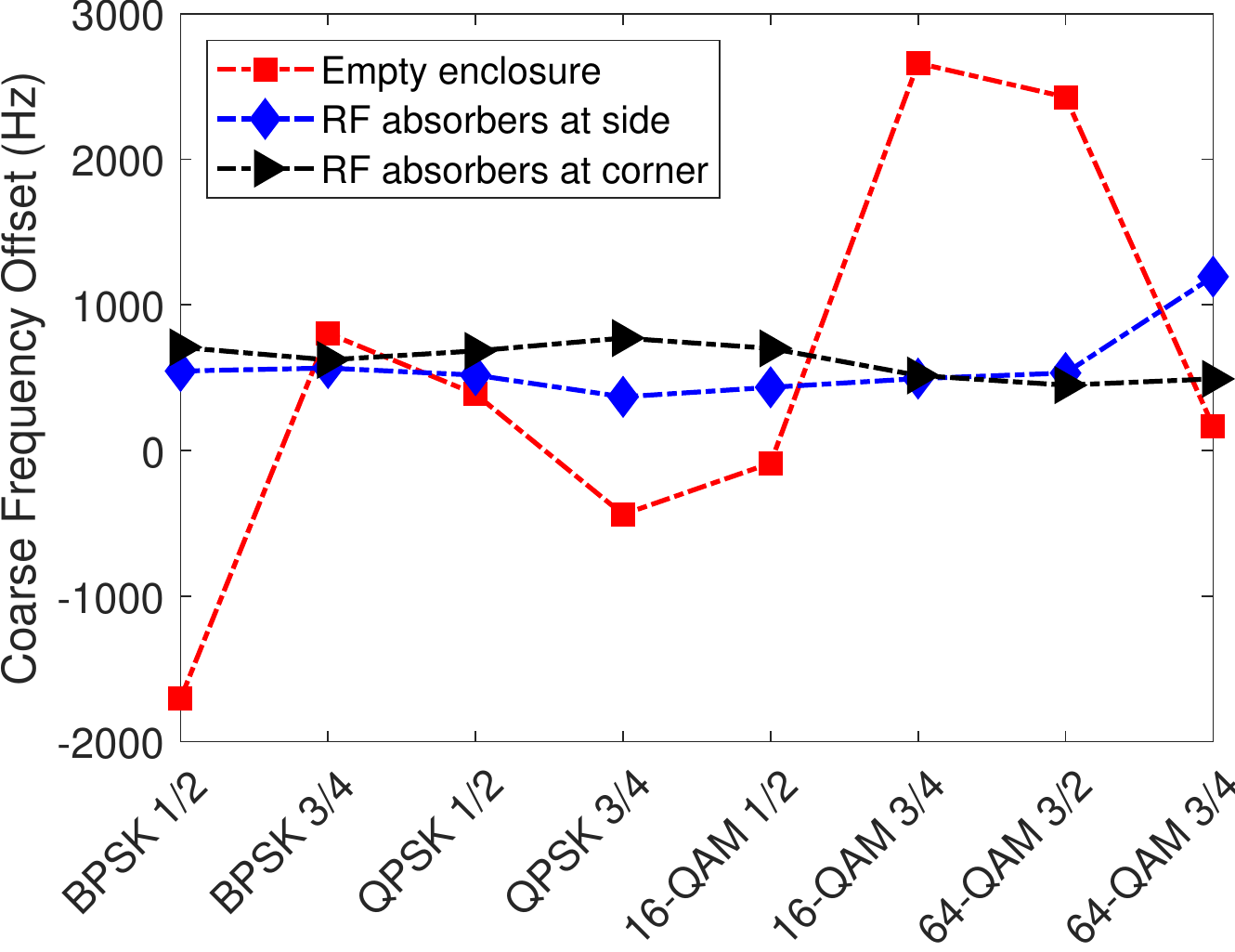}}
    \caption{Coarse frequency offset (a) coarse frequency offset per packet for 64-QAM 3/4 (b) average coarse frequency offset for all MCS in empty enclosure, RF absorbers at side and corner respectively.}
    \label{fig:cfo}
\end{figure}

Fig.~\ref{fig:chan_Est_25mm} shows measured channel state information per subcarrier of twelve OFDM symbols. It can be seen that the effect of frequency selectivity has been reduced after the RF absorber cones are introduced in the metal enclosure. This has the effect of increased coherence bandwidth of the channel. The reduced effect as a result of RF absorber cones can be seen from channel state information and the constellation diagram of modulation technique.Fig.~\ref{fig:cfo} shows coarse carrier frequency offset for 64-QAM 3/4 in empty metal enclosure and average coarse carrier frequency offset for all modulation and coding techniques in empty enclosure and when the RF absorbers are placed at the middle left side and corner of the metal enclosure respectively. It can be seen from Fig.~\ref{fig:avg_cfo} that average coarse carrier frequency offset is higher for empty enclosure in contrast to when metal enclosure is loaded with RF absorbers. Increased coarse carrier frequency offset degrades the performance of signal quality which results in higher RMS and Peak EVM than of when the metal enclosure loaded with RF absorbers.

\begin{figure}[hbt!]
    \centering
    \subfloat[RMS EVM]{\label{subfig:rms_evm}\includegraphics[width=0.7\linewidth]{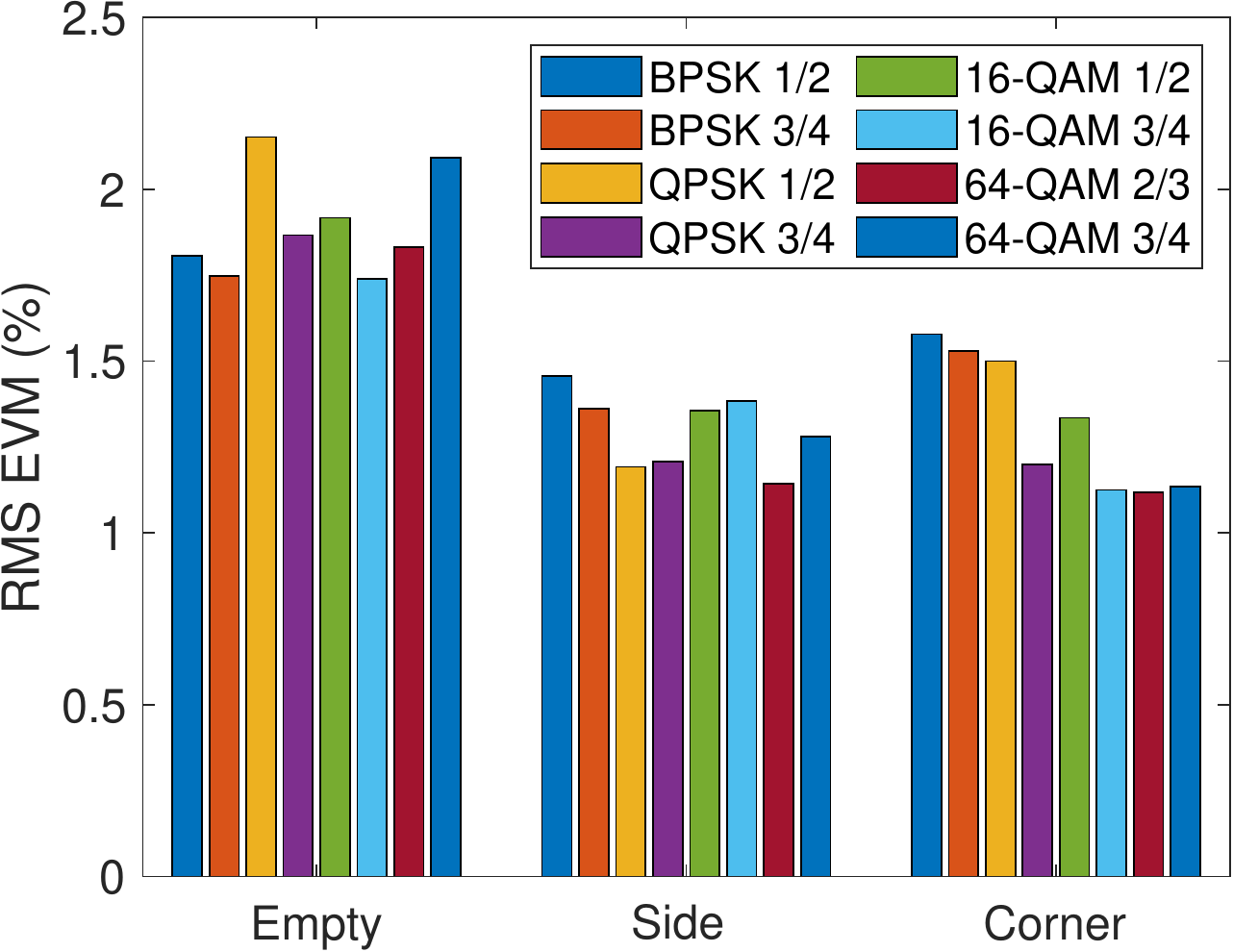}} \hspace{1ex}
    \subfloat[Peak EVM]{\label{peak_evm}\includegraphics[width=0.7\linewidth]{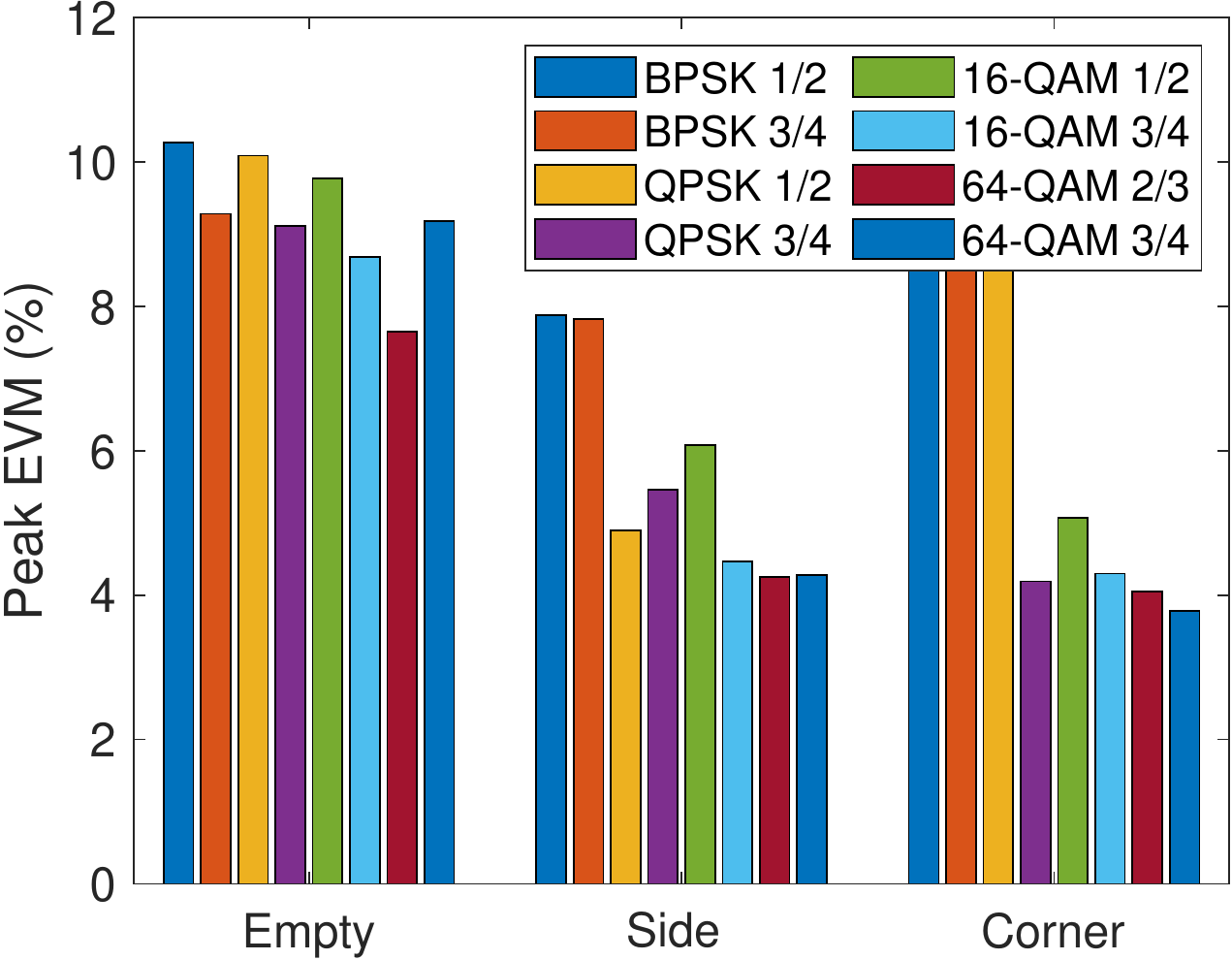}}
    \caption{Near-field RMS EVM and Peak EVM in empty enclosure and when enclosure was loaded with six RF absorber cones. Tx and Rx are separated by a distance of 25mm.}
    \label{fig:nearfield_EVM}
\end{figure}

Fig.~\ref{fig:nearfield_EVM} shows near-field RMS and peak EVM in the empty metal enclosure and in the presence of a loss in the enclosure. It can be seen that RMS and peak EVM is higher for empty enclosure when we have rich scattering. The RMS and peak EVM has reduced when the loss is added in the enclosure. The prime reason beyond this reduction is an increase in coherence bandwidth. It's also interesting to note that peak EVM is higher for lower-order modulation techniques such as BPSK and QPSK than the higher-order modulation techniques. This was also observed in the constellation diagram of BPSK and QPSK modulation techniques.
\begin{figure}[H]
    \centering
    \includegraphics[width=\linewidth]{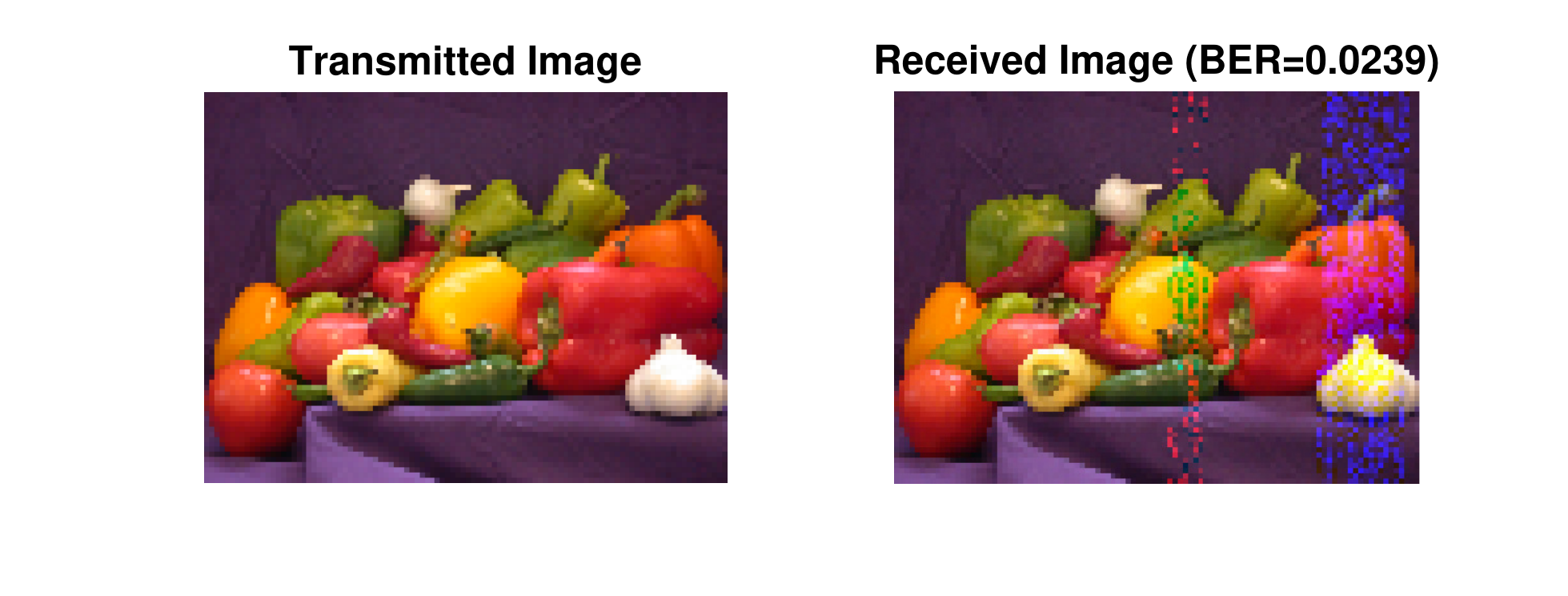}
    \caption{Transmitted and received image for 64-QAM rate 3/4 modulation techniques}
    \label{fig:received_image}
\end{figure}
Fig.\ref{fig:received_image} shows transmitted and received image for 64-QAM 3/4 at distance of 25 mm. For most of modulation techniques at different Tx-Rx distance the received image was perfectly recovered at all loading conditions. However, there were a few instants of received image with errors.
\vspace{-0.45cm}
\section{Conclusion and Future Work}
Near-field measurements were performed without losses and with losses in the middle left side of the metal enclosure and in the corner of the metal enclosure. It was noticed that there was a performance increase in near-field communication with the addition of losses, however, the performance was better when the losses were present in the corner of the metal enclosure.
There were fewer instants of being L-SIG not decoded. Constellation diagrams for all modulation techniques were clean and very few instants of timing recovery failure. Frequency selectivity was removed by adding losses in the metal enclosure.BPSK and QPSK modulation techniques were more prone to timing failures. In IEEE 802.11a the packet is discarded when the header of the packet is corrupted. If the header is not corrupted then the receiver payload bits are compared with the transmitted bits and BER is measured. Future aspect of this research is to check near-field performance and EVM measurements in an anechoic chamber and with slow attack AGC configuration in stationary environment.

\bibliographystyle{IEEEtran}

\bibliography{nf_image.bib} 
\end{document}